# Tuning for Tissue Image Segmentation Workflows for Accuracy and Performance

Corresponding Author:
George Teodoro
teodoro@unb.br
Tel: +55-61-3107-3663;
Fax: +55-61-3107-3663;

Authors:
Luis F. R. Taveira[1], Tahsin Kurc[3,4], Alba C. M. A. Melo[1], Jun Kong[2,5,6],
Erich Bremer[3], Joel H. Saltz[3], George Teodoro[1,3]

[1] Department of Computer Science, University of Brasília, Brasília, Brazil
[2] Department of Biomedical Informatics, Emory University School of Medicine, Atlanta, USA
[3] Department of Biomedical Informatics Department, Stony Brook University, Stony Brook, USA
[4] Scientific Data Group, Oak Ridge National Laboratory, Oak Ridge, USA
[5] Department of Biomedical Engineering, Emory- Georgia Institute of Technology, Atlanta, USA
[6] Department of Mathematics and Statistics, Georgia State University, Atlanta, USA

**ABSTRACT**

We propose a software platform that integrates methods and tools for multi-objective parameter auto-tuning in tissue image segmentation workflows. The goal of our work is to provide an approach for improving the accuracy of nucleus/cell segmentation pipelines by tuning their input parameters. The shape, size and texture features of nuclei in tissue are important biomarkers for disease prognosis, and accurate computation of these features depends on accurate delineation of boundaries of nuclei. Input parameters in many nucleus segmentation workflows affect segmentation accuracy and have to be tuned for optimal performance. This is a time-consuming and computationally expensive process; automating this step facilitates more robust image segmentation workflows and enables more efficient application of image analysis in large image datasets. Our software platform adjusts the parameters of a nuclear segmentation algorithm to maximize the quality of image segmentation results while minimizing the execution time. It implements several optimization methods to search the parameter space efficiently. In addition, the methodology is developed to execute on high performance computing systems to reduce the execution time of the parameter tuning phase. These capabilities are packaged in a Docker container for easy deployment and can be used through a friendly interface extension in 3D Slicer. Our results using three real-world image segmentation workflows demonstrate that the proposed solution is able to (1) search a small fraction (about 100 points) of the parameter space, which contains billions to trillions of points, and improve the quality of segmentation output by 1.20×, 1.29×, and 1.29×, on average; (2) decrease the execution time of a segmentation workflow by up to 11.79× while improving output quality; and (3) effectively use parallel systems to accelerate parameter tuning and segmentation phases.

1. Introduction

We propose and experimentally evaluate a software platform that integrates a suite of methods and tools to enable automatic parameter tuning in analysis algorithms that segment nuclei in digitized images of tissue specimens fixed on glass slides, also called whole slide tissue images (WSIs). Microscopic examination of whole slide tissue specimens by pathologists has long been considered a de facto standard for disease diagnosis and prognosis. Diseased tissue shows changes in tissue morphology, which are indicators of disease onset and progress and provide rich information with which to study disease biology at the sub-cellular level. Manual examination of tissue specimens, however, has had limited use in biomedical research because it is a labor intensive and time consuming process. Advances in digital microscopy scanners have made it possible to capture tissue images at very high resolutions; state-of-the-art scanners can capture images at 100,000x100,000 square pixel resolutions and can automatically scan hundreds of tissue slides rapidly thanks to sophisticated auto-focusing mechanisms. Whole slide tissue images enable quantitative and reproducible analysis of tissue morphology -- the importance of improving precision and reducing inter-observer variability in Pathology studies is well recognized[1-14]. In addition, the Food and Drug Administration (FDA) has recently approved the use of digitized tissue images for diagnostic purposes, both recognizing the value of whole slide tissue imaging in clinical settings and paving the way for routine use of tissue imaging, which we expect will lead to significant increases in the number and volume of WSI datasets for imaging studies. A number of projects have developed tissue image analysis methods[15-21] and shown that quantitative image characterizations from Pathology images can be used to predict outcome and treatment response[16, 22-26]. Nevertheless, development of robust and efficient computerized image analysis workflows to reliably extract imaging features from WSIs remains an open challenge.

Our work targets nucleus segmentation workflows as key part of this open challenge. Segmentation of nuclei is one of the most common steps in WSI analysis, because disease often manifests itself as changes to the properties, such as shape and texture, and organization of nuclei in tissue. A nuclear segmentation workflow detects nuclei and delineates their boundaries. Shape, size, intensity and texture features are computed for each segmented nucleus based on the characteristics of the image (i.e., tissue) within the boundary of said nucleus. These features can then be used to classify images and patients in downstream analyses[27-31]. Thus, the segmentation quality in an image may significantly impact the accuracy and robustness of results obtained from image analysis studies. The inherent complexity of tissue makes it a challenging task to develop accurate and reliable segmentation algorithms. Moreover, many segmentation workflows are configured and controlled by multiple input parameters, which have to be tuned in order to optimize segmentation quality for a given dataset. The parameters oftentimes have to be re-tuned when the segmentation workflow is to be used for a new set of images. This problem is referred to in this work as *the problem of parameter tuning*, i.e., the problem of finding a set of parameter values that generate accurate segmentation results for a set of images.

Our work is motivated by the fact that manual parameter tuning is very time consuming and error-prone, particularly in the context of WSI analysis[32, 33]. An alternative approach is to manually segment several image tiles accurately, generating a ground truth segmentation set for a given image dataset, and then apply a computerized method to search for a set of parameter values that produce the best segmentation output with respect to the ground truth. This also is a challenging task, because parameter search space for a segmentation workflow can be very large, containing billions or trillions of points as is shown in Table 1 for the example segmentation workflows studied in this work. Moreover, the computational cost of evaluating a single point in the parameter space, which involves segmenting an image tile and computing a quality metric for the segmentation results, can be very high. The parameter tuning process becomes even more challenging when it involves multiple conflicting objectives such as the quality of segmentation output and the execution time of the segmentation process[34-40].

Table 1. Input parameter sets of three example segmentation wokflows.

| Parameter | Description | Range Value |
|---|---|---|
| B/G/R | Background detection thresholds | [210, 220,…, 240] |
| T1/T2 | Red blood cell thresholds | [2.5, 3.0,…, 7.5] |
| G1/G2 | Thresholds to identify candidate nuclei | [5, 10,…, 80] [2, 4,…, 40] |
| MinSize | Area threshold of candidate nuclei | [2, 4,…, 40] |
| MaxSize | Area threshold of candidate nuclei | [900,…, 1500] |
| MinSizePl | Area threshold before watershed | [5, 10,…, 80] |
| MinSizeSeg | Area threshold from final segmentation | [2, 4,…, 40] |
| MaxSizeSeg | Area threshold from final segmentation | [900,…, 1500] |
| FillHoles | Propagation neighborhood | [4-conn, 8-conn] |
| MorphRecon | Propagation neighborhood | [4-conn, 8-conn] |
| Watershed | Propagation neighborhood | [4-conn, 8-conn] |

(a) Parameters of the Morphological Operations and Watershed based segmentation pipeline. The search space contains about 21 trillion parameter points.

| Parameter | Description | Range Value |
|---|---|---|
| OTSU | OTSU threshold value | [0.3, 0.4,…, 1.3] |
| Curvature Weight | Curvature weight (CW) for the level-set | [0.0, 0.05,…, 1.0] |
| MinSize | Area threshold for nuclei | [1, 2,…, 20] |
| MaxSize | Area threshold for nuclei | [50, 55, …, 400] |
| MsKernel | Radius in Mean-Shift calculation | [5, 6,…, 30] |
| LevelSetIt | Number of iterations of the level set computation | [5, 6,…, 150] |

(b) Parameters of the Level-Set and Mean Shift based segmentation pipeline. The search space contains about 1.4 billion parameter points.

| Parameter | Description | Range Value |
|---|---|---|
| OTSU | OTSU threshold value | [0.3, 0.4,…, 1.3] |
| Curvature Weight | Curvature weight (CW) for the level-set | [0.0, 0.05,…, 1.0] |
| MinSize | Area threshold for nuclei | [1, 2,…, 20] |
| MaxSize | Area threshold for nuclei | [50, 55, …, 400] |
| Watershed | Propagation neighborhood | [4-conn, 8-conn] |
| LevelSetIt | Number of iterations of the level set computation | [5, 6,…, 150] |

(c) Parameters of the Level-Set and Watershed based segmentation pipeline. The search space contains about 96 million parameter points.

The problem of parameter tuning and optimization has been investigated in several projects [34, 41-52]. Majority of the previous works use solutions for particular segmentation models. A pseudo-likelihood is used in[49] to estimate parameters for a conditional random field based algorithm. Graph cuts are employed to compute maximum margin efficient learning in segmentation parameters[50]. Open-Box[52] is another interesting solution specific to segmentation algorithms based on spectral clustering. It deals with the optimization by exposing key components of the segmentation to the user. The Tuner system[32] treats a segmentation pipeline as a black-box process, but it uses statistical models to explore the parameter space. Our work has several novel improvements over the prior art. Our approach tunes a segmentation workflow as a black-box with efficient optimization algorithms that quickly converge to desired results. It also allows for the use of multiple auto-tuning algorithms and multiple objectives as well as several domain specific metrics for evaluating algorithm output. It is implemented to take advantage of high performance computing (HPC) systems to accelerate runs in the parameter-tuning phase. Another related work proposed the use of parameter auto-tuning and efficient parameter sensitivity analysis[47] in microscopy image with single objective optimization. We extend the previous work to develop a multi-objective parameter tuning software platform. Our contributions can be summarized as follows:

1. We have adapted multi-objective optimization algorithms, some of which were developed to optimize performance of other classes of applications, in pathology image analysis and implemented them in an integrated platform. As such, our goal is to demonstrate that the target class of applications can substantially benefit from these methods and to evaluate the effectiveness of efficient algorithms in the domain and, for instance, provide a reference and understand their performance.

2. We show via experimental evaluation that pathology image analysis can substantially benefit from multi-objective parameter optimization that targets the quality of analysis output and the execution time of analysis. The proposed platform supports two objectives in parameter optimization; reducing the execution time of an image segmentation run and increasing the quality of the segmentation results. These objectives conflict with each other in that reducing execution time often leads to reduction in segmentation quality, and vice versa. Our approach identifies parameters that produce segmentation results with good quality and reduced segmentation execution time. Users can adjust the weights of the objectives (quality vs execution time) to tune the segmentation pipeline according to their priorities. Our experimental evaluation with three nucleus segmentation workflows shows very good results. First, the quality of segmentation can be improved up to 7.8× when only one objective is set. Second, the multi-objective tuning can speed up the segmentation process by 11.79× while improving the segmentation quality by 1.28× (please see Table 2).

3. We package our implementation as a Docker container with a RESTFul interface for easy deployment as a service. A user can run the parameter auto-tuning infrastructure as a service on a local machine or as a remote server in a Cloud environment. A Cloud deployment can benefit scientists without access to high performance systems. Our implementation supports a variety of spatial queries that include spatial cross-matching, overlap and spatial proximity computations, used to derive segmentation quality metrics as Dice and Jaccard coefficients[53].

4. We integrate the Docker container with 3D Slicer[54] as part of the SlicerPathology extension[55, 56]. This integration provides a graphical user interface for a researcher to interact with the infrastructure.

5. We propose a high performance computing approach that integrates parameter auto-tuning processes and spatial query capabilities for comparing analysis results in order to reduce the execution time of auto-tuning.

## 2. Materials and Methods

An overview of the auto-tuning platform is presented in Figure 1. Using the 3D Slicer Pathology extension from 3D Slicer, a user specifies input images and corresponding segmentation masks, the set of parameters to be tuned and their value ranges, and the optimization algorithm to be used. In step 2 in

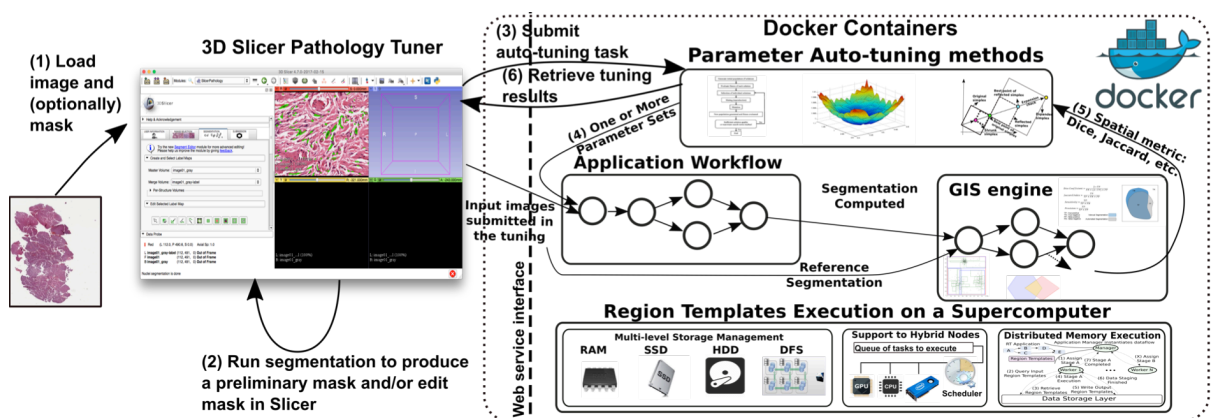

Figure 1. Auto-tuning framework.

Figure 1, the user can employ the SlicerPathology module [37] in 3D Slicer [36] to create a segmentation mask for an input image. The user then invokes the auto-tuning infrastructure to execute the auto-tuning task – we have implemented a prototype extension to SlicerPathology so a user can submit the auto-tuning task through the SlicerPathology graphical user interface. The task is received in the infrastructure through a web-service interface and executed as the tuning system becomes available. During the tuning, the optimization algorithm selects parameter sets and the segmentation workflow is executed for each of the sets. The output of an execution are the segmentation results (image masks), which are compared to ground truth results using spatial metrics (Dice and Jaccard coefficients)[53]. The value of the computed metric is used as input to the auto-tuning or optimization algorithm to guide the search. This search and comparison loop is repeated until either the desired objective is met or the maximum number of iterations is reached. In the rest of this section, we describe the multi-objective strategy used to tune the applications execution times and results quality, the optimization algorithms employed and evaluated in this work, the application workflows used in the evaluation, and the implementation details of the auto-tuning framework.

## 2.1. Multi-objective Auto-tuning Methodology

Multi-objective tuning deals with optimization problems with conflicting goals, e.g., result quality and execution time or performance and resource usage. There are multiple solutions to a multi-objective optimization problem. Existing works in the literature typically employ one of three fundamental approaches[35, 37]: (i) *A posteriori* in which the largest number of possible solutions is first sought and then the one that best fits the problem is selected; (ii) *A priori insertion*, in which there exists a preference about the type of solution most appropriate to the problem at hand. The search is directed to find that type of solutions; and (iii) *Progressive insertion of preferences* that is done by targeting the choices of a Decision Maker (a person skilled in the problem domain) during optimization to regions that are more likely to contain appropriate solutions.

Our work employs an *a priori insertion* approach, because (a) in our problem, it is not possible to test a large number of parameter's combinations as in the *a posteriori*. Calculating a large number of combinations (i.e., the set of Pareto optimal solutions or those in which none of the objectives can be improved without affecting another objective[34, 39]) would be very expensive due to the high cost of the test or fitness function (segmentation)[37]; (b) the use of *progressive insertion of preferences* during the search would require the intervention of a domain expert, but we want to carry out the optimization process automatically and minimize the users' burden.

We have chosen the Scalarization approach[37, 38, 57] for inserting a priori preferences and combining objectives in an optimization function. This approach can efficiently solve multi-objective optimization problems through adaptation of single-objective optimization methods. There exists a large class of optimization methods that efficiently solve single-objective problems and can be used with Scalarization. Moreover, in our case, we are typically interested in a given set of weights or preference selected by a user according to the objective of the optimization. We use a linear Scalarization that assigns weights ($w_i$) to each objective ($f_i(x)$), so that the sum of the weights is equals to 1. For N objectives, the function to be optimized is as follows: $f(x) = \sum_{i=1}^{N} w_i \times f_i(x)$. In our evaluation, we mainly target the simultaneous optimization of a segmentation quality/accuracy that is maximized along with the minimization of the segmentation workflow execution time.

## 2.2. Optimization Algorithms

The optimizations methods implemented in our work include Nelder-Mead simplex (NM)[58], Parallel Rank Order (PRO)[41], Bayesian optimization algorithm (BOA)[44], and Genetic Algorithm (GA)[37, 42]. Briefly, the NM is a commonly used optimization algorithm in multidimensional space problems in which the derivatives are may not be known. It is a heuristic search method that explores the search space using a simplex or a special polytope with k +1 vertices, where k is the dimensionality of the search space. The search is carried out by modifying and moving the simplex through a set of complementary operations, such as reflection expansion, contraction, and shrink, that are intended to either quickly find a minimum in the region being explored or to leave a local minima region. The PRO is similar to NM, meaning that it uses the same searching mechanisms and operations, but it enables the evaluation of multiple points of the simplex concurrently. The GA optimization algorithm models the auto-tuning problem with individuals whose genes represent the application parameters. In our GA algorithm, the first set of individuals is randomly initialized, whereas they are modified or evolved among iterations of the algorithm using crossover and mutation. The crossover uses one-point crossover in which individuals' parents are combined by swapping parts of their genes starting into a single point. The crossover between pairs of individuals occurs with a probability of C. After that transformation, the mutation in each gene can take place with a probability of M. Once the new population is created, it is evaluated via segmentation workflow runs (fitness function), and the results are used again to create another generation of individuals. The tuning with GA executes for a number of iterations selected by the end-user. The probabilities C and M were selected experimentally as 0.5 and 0.3, respectively, as those are the values that maximize the GA performance in our case.

The BOA[44] is an iterative process that develops a global statistical model of the objective function. This probabilistic model is exploited to make decisions about the next point in the search space for which the objective function should be evaluated. It uses information from the model or previous runs in this decision and minimizes the number of function evaluations. As such, this method is expected to be competitive[59] for objective functions whose evaluations are costly, which is the case of medical image analysis.

### 2.3. Segmentation Workflows

We have evaluated our approaches using three segmentation workflows presented in Figure 2. We include these workflows as part of our software distribution. The analysis workflows of tissue images used in our studies compute information from the images that include segmented objects (e.g., nuclei or cells) and about 30-50 features per object (shape, intensity, and texture features). The overall image computation workflow includes Normalization, Segmentation, Feature computation, and other data analysis stages, being the first three the most expensive. In this work, we focus on the study of the Segmentation stage. Figure 2 presents three analysis workflows used in this work. The workflows have the same high-level structure, but they are different with respect to the approaches used to implement the Segmentation stage. The first workflow (Figure 2 (a)) uses morphological operations and Watershed in the segmentation[60], whereas the second one (Figure 2 (b)) performs the segmentation based on Level Set and Mean-Shift clustering[55]. The third workflow uses Level Set and Watershed for declumpling[55] (Figure 2 (c)). The operations within these workflows are shown in the figures. Please see Table 1 for the list of parameters for the Segmentation phase of each workflow.

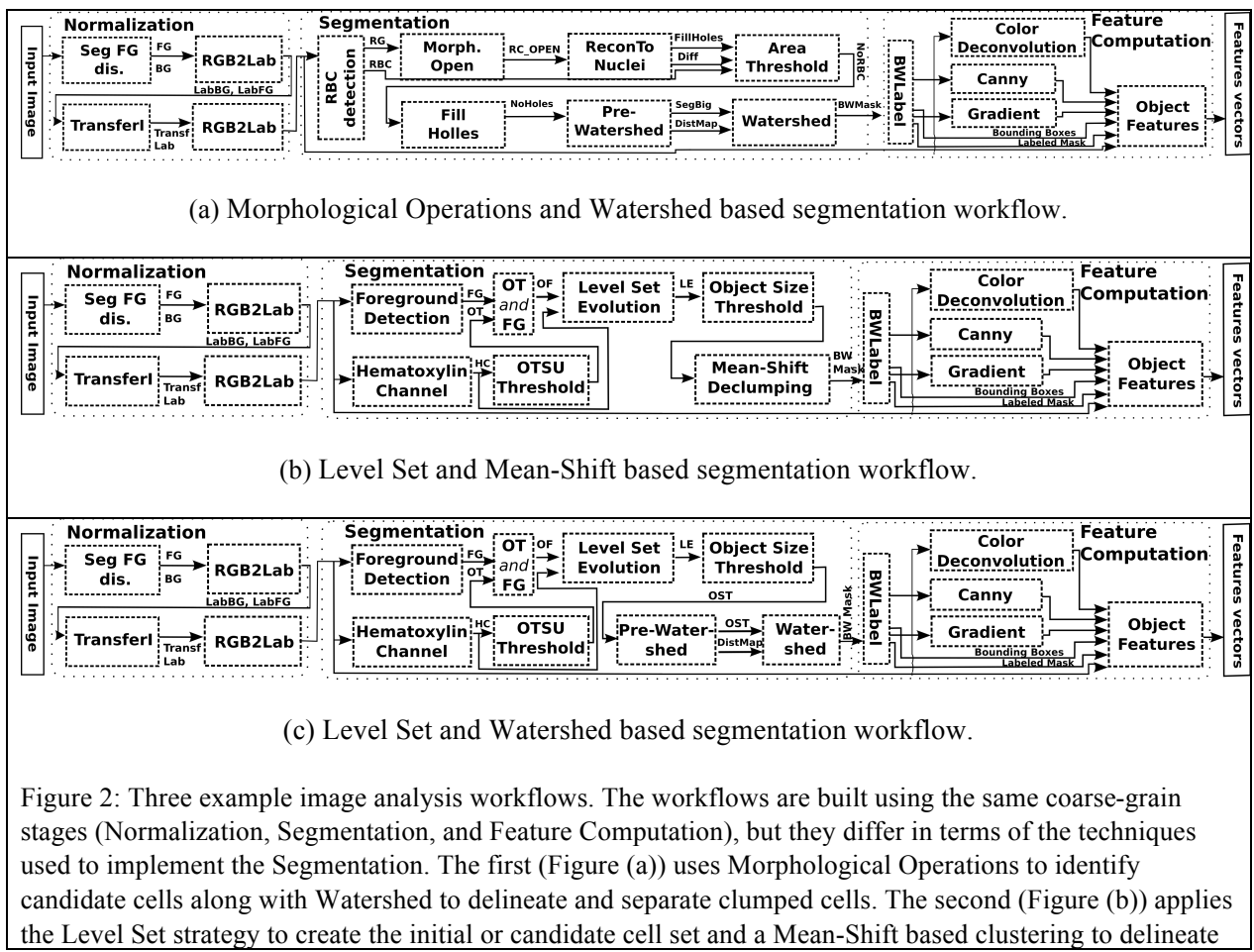

(a) Morphological Operations and Watershed based segmentation workflow.

(b) Level Set and Mean-Shift based segmentation workflow.

(c) Level Set and Watershed based segmentation workflow.

Figure 2: Three example image analysis workflows. The workflows are built using the same coarse-grain stages (Normalization, Segmentation, and Feature Computation), but they differ in terms of the techniques used to implement the Segmentation. The first (Figure (a)) uses Morphological Operations to identify candidate cells along with Watershed to delineate and separate clumped cells. The second (Figure (b)) applies the Level Set strategy to create the initial or candidate cell set and a Mean-Shift based clustering to delineate

> and separate clumped cells. The third (Figure (c)) uses the Level Set to create candidate cells and the Watershed to separate clumped ones.

## 2.4. Software Implementation

This section presents the implementation aspects of the main modules of our auto-tuning platform presented in Figure 1. First, in Section 2.4.1, we describe the Region Templates framework that is used to implement the applications for efficient execution on distributed high-performance computing systems, and is the baseline solution in which the tuning methods and spatial comparison engine were integrated. The spatial comparison engine that computes the differences between different segmentation workflows is details in Section 2.4.2. The containerization and integration of our solutions with 3D slicer for simplifying the system deployment and interaction with the proposed tuning tools are then discussed in Section 2.4.3.

### 2.4.1. Execution on High-Performance Machines with Region Templates

The auto-tuning methods are deployed in the Region Templates (RT)[61] for efficient execution of image analysis pipelines on parallel machines. The integration of applications or workflows with RT for tuning is performed using an interface in which the user exports the parameters to be tuned and their value ranges. In the same interface the user can also choose the optimization algorithm to be used and modify the weights given to each parameter in multi-objective tuning tasks.

We implemented the three example segmentation pipelines in RT to tune and accelerate their execution. A RT application stage itself can be composed of lower-level operations organized into another dataflow, and different scheduling strategies may be used at each level. The runtime system implements a Manager-Worker execution model for assignment of work among nodes of a distributed memory machine. The application Manager creates instances of (coarse-grain) stages, and exports the dependencies between them. The assignment of work from the Manager to Worker nodes is performed at the granularity of a stage instance using a demand-driven mechanism.

Each Worker uses multiple computing devices in a node by dispatching fine-grain tasks (operations that implement a stage instance) for execution in a CPU core or a co-processor (e.g., Intel Phi or GPU). Different scheduling strategies and runtime optimizations were developed targeting heterogeneous computing devices[62-66]. Further, RT also implements optimizations to reduce the cost of exchanging data among stages and to improve data access locality are also implemented and available to the applications transparently[67].

### 2.4.2. Spatial Query Module for Computing Error Metrics

The computation of quality (or error) metrics to guide the parameter tuning process involves spatial queries and comparison operations. We implemented a spatial query module, called RT GIS Engine, to speed up the quantitative comparison of segmentation results via a query-based interface with which queries are expressed using a SQL-like language. The implementation uses a query engine[68] that supports several spatial operations, including spatial cross-matching, overlay of objects, spatial proximity computations between objects, and global spatial pattern discoveries. These operations are used to compute high-level metrics for comparison of results from different analysis runs. The quality metrics include Dice Coefficient, Jaccard Coefficient, Intersection Overlapping Area and Non-Overlapping Area[53].

The workflow for computation of the comparison and error metrics is depicted in Figure 3. The user application computes a mask and passes it along with the reference mask as input to the RT GIS engine. In order to execute spatial queries, the objects (i.e., cell nuclei) identified in the masks are converted into polygons and processed into the query engine. Each of the metrics implemented uses a set of queries (expressed using a SQL-like language) that are combined to compute the user-selected metric. The spatial queries are executed using a Hilbert R*-Tree index[69] to quickly identify intersecting objects and minimize the computation costs. First, the R*- Trees are built from objects minimum bounding boxes in each mask (computed and reference). The spatial filtering operation is performed to identify possibly overlapping objects (those with intersecting bounding boxes), which are refined to those that overlap. This set is passed to a final phase that computes the spatial measurements. The entire query engine is deployed as a generic workflow stage in RT. As such, the query engine can execute on parallel machines and may have several copies running in the computing environment as a regular RT application stage.

### 2.4.3. Containerization and Integration with 3D Slicer

In order to simplify the use the auto-tuning methods and the execution platform proposed in this work, we have (i) packaged our implementation as a service in a Docker container[70] and (ii) exported the main functionalities through a user-friendly interface with the Pathology module[37] of 3D Slicer[36]. With the Docker container of the auto-tuning platform, the user can easily build the entire system and deploy it into local or remote computing systems (i.e., Cloud providers), whereas the Pathology module contains a graphical interface for interacting with the entire system.

The auto-tuning process is invoked by calling the service hosted in the Docker container. The user specifies the input image, the reference mask, a set of parameters to be tuned, an optimization algorithm, and a quality/error metric. This request is sent to the service via a RESTFul interface call. The service will parse the input request and insert it into a queue of tuning requests. A tasks hander is returned by the services and can be used by the client application (Pathology Slicer in our case) for querying the status or results of the given tuning request. As such, the client side of the application or Slicer is not blocked during the execution of the auto-tuning. The queue of requests is handled internally by the web-service running into the container. In our current implementation, the web-service processes those requests by executing multiple instances of the RT implementation of the application being tuned in a demand-driven basis as the computing resources (CPU cores or GPUs) become available. Once the request is processed, it is placed into a queue of completed requests and remains available for the user to retrieve the results. The auto-tuning execution itself goes through a set of steps that consists of executing the segmentation workflow, computing the quality metric, and computing a new set of parameters to evaluate, as described in Section 2. This process is executed until it converges. Another aspect important aspect that we want to highlight is that the web-service interface is built independently of the 3D Slicer. As such, the same interface could be used to integrate with other tools.

Our Slicer module interface provides a set of additional features that can speed up and simplify the tuning. For instance, it enables the integration with remote repositories[1] for loading the data employed in the auto-tuning process, which alleviates the burden on the user with respect to the data management. Additionally, tools available in 3D Slicer module allow for the user to delineate objects and create reference (ground truth) masks manually. This process may also be sped up by starting the ground-truth generation with a mask produced by the segmentation workflow and parameter values chosen by the scientist (Step 2 in Figure 1). The results of this segmentation are presented in the Slicer, and can be corrected/modified manually using the editing tools to change the polygons that describe objects found in the segmentation, instead starting the mask generation from scratch.

## 3. Results and Discussion

### 3.1. Experiment Setup

The experiments were conducted on the Stampede distributed memory machine. Each node on Stampede has dual Intel Xeon E5-2680 processors, an Intel Xeon Phi SE10P co-processor and 32GB RAM. The nodes are inter-connected via Mellanox FDR Infiniband switches. The auto-tuning methods and the example segmentation pipelines are implemented in the Region Templates (RT) framework[61] for efficient execution on this machine. In this implementation, input images are partitioned into image tiles – each image tile can be processed independent of other tiles for nucleus segmentation.

In the experiments the genetic algorithm (GA) method was configured to evolve 10 individuals for 10 generations and set up with a mutation rate of M=0.3 and a crossover rate of C=0.5, because this setup

---

[1] http::/quip1.bmi.stonybrook.edu, for example, contains thousands of images from TCGA

experimentally led to best results. The NM, PRO, and BOA algorithms were configured to stop after testing 100 points in the search space, which asserts that all optimization algorithms perform 100 application runs. We have repeated all experiments 10 times. Average standard deviation is smaller than 1% for the Watershed based workflows and 3% for the Mean-Shift workflow. The time to select the next set of parameters varies among the optimization methods. It was about 77s for BOA and around 10ms for the other algorithms. This cost is amortized by the high execution times of the segmentation. The quality of segmentation results was quantified with the *Average Dice Coefficient*, which ranges from 0.0 to 1.0, in which higher values mean a better agreement with ground-truth segmentation. The experiments used 15 image tiles extracted from Glioblastoma multiforme(GBM) WSIs and manually segmented by a pathologist.

### 3.2. Multi-objective Parameter Auto-Tuning: Segmentation Quality and Execution Time

This section evaluates our methodology to maximize the quality of segmentation results and to minimize the execution time of the 15 images used in our analysis. The objectives are combined into a single optimization function using Scalarization, and the user defines the weight for each objective. To simplify the weighting, we have normalized execution times between 0 and 1 (higher is better), which is the same range of the Average Dice Coefficient used for quality.

Table 2. Results for the multi-objective auto-tuning as compared to the application default parameters. The experiments were carried out using all supported optimization algorithms as the weights for the results quality (Average Dice) and execution times were varied. The results are presented as an average for the 15 input example images, and the best results for the multi-objective function are highlighted in each case.

| Segmentation Algorithm | Weights | | Average Dice | | | | | | Speedup vs. Default | | | |
|---|---|---|---|---|---|---|---|---|---|---|---|---|
| | Metric | Time | Default | GA | NM | PRO | BOA | Improv. | GA | NM | PRO | BOA |
| Morphological Operations + Watershed | 1 | 0 | 0.65 | 0.77 | 0.76 | 0.76 | **0.78** | 1.20 | - | - | - | - |
| | 1/2 | 1/2 | 0.65 | 0.70 | 0.72 | **0.73** | 0.70 | 1.10 | 1.09 | 1.08 | **1.07** | 1.07 |
| | 2/3 | 1/3 | 0.65 | **0.74** | 0.74 | 0.74 | 0.72 | 1.13 | **1.08** | 1.07 | 1.06 | 1.06 |
| | 4/5 | 1/5 | 0.65 | **0.75** | 0.75 | 0.74 | 0.75 | 1.14 | **1.07** | 1.07 | 1.05 | 1.04 |
| Level Set + Mean-Shift Declumping | 1 | 0 | 0.61 | **0.79** | 0.75 | 0.73 | 0.78 | 1.29 | - | - | - | - |
| | 1/2 | 1/2 | 0.61 | 0.70 | **0.66** | 0.63 | 0.71 | 1.14 | 3.99 | **4.91** | 5.09 | 0.67 |
| | 2/3 | 1/3 | 0.61 | 0.73 | **0.69** | 0.68 | 0.72 | 1.19 | 2.41 | **3.74** | 2.85 | 0.61 |
| | 4/5 | 1/5 | 0.61 | 0.74 | **0.71** | 0.71 | 0.75 | 1.21 | 1.68 | **2.61** | 2.38 | 0.48 |
| Level Set + Watershed Declumping | 1 | 0 | 0.61 | **0.79** | 0.78 | 0.75 | 0.78 | 1.29 | - | - | - | - |
| | 1/2 | 1/2 | 0.61 | 0.74 | **0.69** | 0.69 | 0.73 | 1.13 | 14.26 | **16.05** | 13.01 | 1.38 |
| | 2/3 | 1/3 | 0.61 | **0.77** | 0.73 | 0.72 | 0.72 | 1.26 | **12.68** | 13.99 | 11.85 | 1.42 |
| | 4/5 | 1/5 | 0.61 | **0.78** | 0.75 | 0.71 | 0.76 | 1.28 | **11.79** | 11.61 | 10.96 | 1.27 |

Experimental results with multi-objective tuning are presented in Table 2 for the three segmentation workflows. The weights of the objectives are varied to evaluate the ability of the parameter auto-tuning framework to find parameters for different user preferences. The results for single objective tuning (execution time weight set to 0) are also presented for reference. The experimental results for the workflow using Morphological Operations and Watershed show that the optimization algorithms were able to improve the quality of segmentation results by up to 1.14× and speed up execution by 1.07× compared with the segmentation quality and execution times using the default parameters. Moreover, the results show a consistency in segmentation quality improvement when the weight of this component is increased, i.e. higher quality metric weights resulted in better Dice values. Also, GA and NM achieved a slightly better performance than the other methods.

Results for the Level Set based workflows are also presented in Table 2. As is shown in the table, NM and GA attained the best-aggregated multi-objective optimization value (marked in bold font) when segmentation quality and execution times are considered in all configurations. When the workflow used the Watershed declumping method, GA was able to find parameter sets with which the quality metric was increased by 1.28× and the execution time reduced by 11.79× compared with the default parameters. The parameter sets found for the workflow with Level Set and Mean-Shift are also very good. The segmentation results for this workflow can be significantly improved while at the same time the execution time is reduced. When the segmentation operations are applied in datasets with thousands of WSIs, these improvements can translate to significant reduction in resource usage, much faster analysis of data, and ability for large-scale studies.

The PRO optimization method also was able to find parameter configurations with which the execution times of the workflows significantly improved. However, the reductions in execution times were achieved with a higher penalty in the quality metric than what GA achieved. The BOA method, on the other hand, was not able to find parameter sets that resulted in similar levels of reduction in the execution times. Indeed, for Level Set with Mean-Shift, parameters selected by BOA resulted in a higher execution time vs. those using the default parameters. These results indicate that BOA is efficient for single-objective tuning, while GA and NM are the best methods for multi-objective tuning.

We also examined why the gains in execution time with the Morphological Operations and Watershed based workflow were smaller than those with the Level Set based segmentations. We found that in practice, variations in input parameters have a small impact in the execution time of the first workflow. As such, it was not a failure of the optimization methods in finding good parameter sets, but a characteristic of the segmentation strategies used.

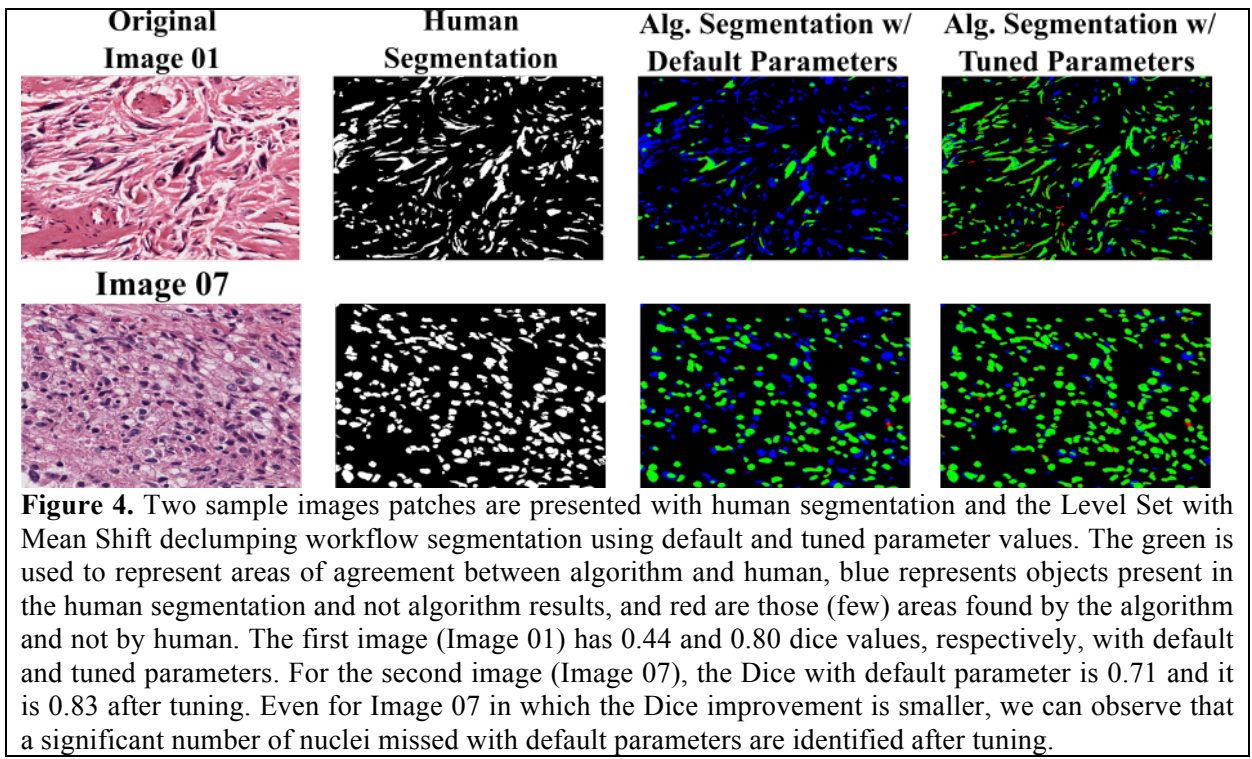

**Figure 4.** Two sample images patches are presented with human segmentation and the Level Set with Mean Shift declumping workflow segmentation using default and tuned parameter values. The green is used to represent areas of agreement between algorithm and human, blue represents objects present in the human segmentation and not algorithm results, and red are those (few) areas found by the algorithm and not by human. The first image (Image 01) has 0.44 and 0.80 dice values, respectively, with default and tuned parameters. For the second image (Image 07), the Dice with default parameter is 0.71 and it is 0.83 after tuning. Even for Image 07 in which the Dice improvement is smaller, we can observe that a significant number of nuclei missed with default parameters are identified after tuning.

Figure 4 illustrates improvements in segmentation results via parameter auto-tuning for two images. The images were segmented with the Level Set and Mean-Shift based workflow. The second column in the figure shows the segmentation results from manual segmentation; the third column shows the results with the default parameters; and the fourth column with the auto-tuned parameter values. The green areas in the images refer to agreement between the segmentation generated by the workflow and the manual segmentation, whereas the blue ones are the areas missed by the segmentation pipeline. The red areas refer to those segmented by the pipeline but not by the pathologist. Please, notice that there are very few red points, which means that the algorithms do not significantly detect objects also not found by human (Please, zoom-in for better visualization).

## 3.3. Cross-Validation

This section computes Monte Carlo cross-validation analysis repeated 10 times that separates the 15 input images into two sets for training and testing. The application parameters are tuned using the training set containing 20% of the images and further evaluated into the remaining 80% images of the testing set. The experiments used the GA optimization and evaluated 100 points. The tuning experiments were repeated 10 times and the standard deviation is smaller than 2% for the Morphological Operations workflow and 6% for the Level Set workflows.

The results are presented in Table 3. Since the results are computed using a random selection of the training and testing sets, results using different weights employ different sets and, as such, default metric values are not the same. However, the same sets are used within each weight combination (table row) for a fair comparison between default and tuned parameters. For the Morphological Operations based workflow, the tuning platform found a set of parameters that improved the quality results in over 1.10× and the segmentation speed up by 1.09× and, similarly to the previous experiments, it finds different trade-offs among segmentation results quality and speed of the execution as weights are varied.

Table 3. One group cross-validation evaluation. Average Dice improvements and execution time reductions for all segmentation workflows in a Monte Carlos cross-validation w/ 10 repetitions.

| Segmentation Algorithm | Weights | | Avergea Dice (0-1) | | Speedup |
|---|---|---|---|---|---|
| | Metric | Time | Default | Tuned (GA) | |
| Morphological Operations + Watershed | 1 | 0 | 0.65 | **0.72** | - |
| | 1 | 1 | 0.66 | **0.68** | 1.09 |
| | 2 | 1 | 0.65 | **0.71** | 1.07 |
| | 4 | 1 | 0.66 | **0.73** | 1.06 |
| Level Set + Mean-Shift Declumping | 1 | 0 | **0.63** | 0.60 | - |
| | 1 | 1 | **0.59** | 0.57 | 3.10 |
| | 2 | 1 | **0.62** | 0.58 | 2.32 |
| | 4 | 1 | **0.62** | 0.60 | 1.64 |
| Level Set + Watershed Declumping | 1 | 0 | **0.60** | **0.60** | - |
| | 1 | 1 | **0.60** | 0.58 | 8.97 |
| | 2 | 1 | **0.59** | **0.59** | 10.50 |
| | 4 | 1 | **0.62** | 0.60 | 7.37 |

For the Level Set workflows, regardless of the declumping method used, the optimization algorithms were not able to find parameter sets that improve the segmentation quality and execution time together. However, the execution times were significantly improved (up to 10.5×) for both declumping cases for all weight combinations with a small decrease in the segmentation quality. The Level Set workflows are very sensitive to the nuclei shape, and parameters used for elongated nuclei, for instance, will not perform well in round ones. This indicates that a single set of parameters optimized for images containing different cells shapes will not be optimal for neither of them. Instead, the algorithm should use different parameter sets according the expected nuclei structure.

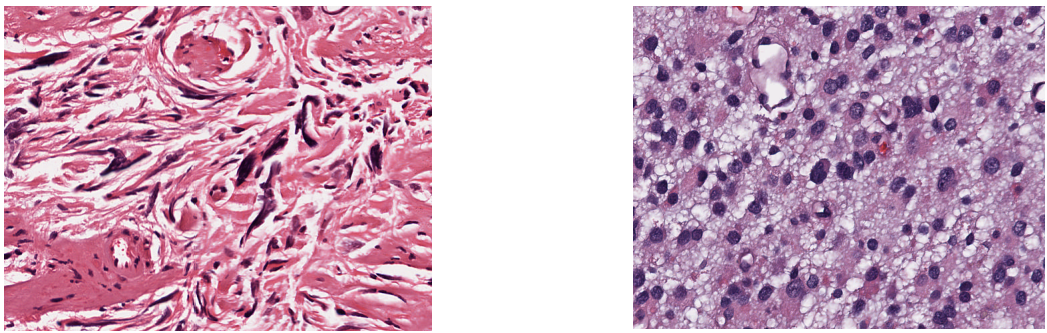

Figure 5. Examples of images in each of the two groups for cross-validation. The first group of images contains those with elongated nuclei (group 1), whereas the second group of images contains more round shaped ones (group 2). The Monte Carlos cross-validation w/ 10 repetitions is the executed in each of the groups isolated.

In order to validate this observation, we performed another cross-validation in which images were separated into two groups: those with more elongated and those with round nuclei. We executed a cross-validation in each of the groups separately. We classified 5 images in the first group and 10 in the second group. Examples of images in these groups are presented in Figure 5. For the first group, 1 image is selected for training and the other 4 images to test, whereas in the second group 2 images are included in the training and the other 8 images in the testing set.

The results are presented in Table 4. The optimization algorithm found parameters that improved the quality and execution time of the workflows for most of the weights. In the workflow with Watershed declumping, the same quality result observed in the single-objective optimization (0.70) was attained, but the workflow was accelerated in 10.25× in the multi-objective case.

Table 4. Two groups cross-validation test. Average metric improvements and execution time reductions for the Level Set based segmentation workflow using the two groups of validation data (4+8 images).

| Segmentation Algorithm | Weights | | Avergea Dice (0-1) | | |
|---|---|---|---|---|---|
| | Metric | Time | Default | Tuned (GA) | Speedup |
| Level Set + Mean-Shift Declumping | 1 | 0 | 0.61 | **0.69** | - |
| | 1 | 1 | 0.62 | **0.62** | 8.56 |
| | 2 | 1 | 0.61 | **0.62** | 1.72 |
| | 4 | 1 | 0.62 | **0.69** | 1.37 |
| Level Set + Watershed Declumping | 1 | 0 | 0.62 | **0.70** | - |
| | 1 | 1 | 0.61 | **0.62** | 14.55 |
| | 2 | 1 | 0.61 | **0.66** | 13.26 |
| | 4 | 1 | 0.61 | **0.70** | 10.25 |

## 4. Conclusions

Pathology image segmentation workflows are sensitive to changes in input parameters, and an input parameter configuration that performs well with a set of images may not compute good segmentation results, for instance, for another dataset. Tuning the application parameters is important to maximize the quality of the results and/or reduce the application's execution time. The main challenges with tuning includes (i) the large number of parameter combinations; (ii) the high cost of evaluating a point in the search space due compute expensive nature of the segmentation workflow; (iii) the difficulty of manually evaluating the search space and evaluate the quality of a segmentation result.

In order to address these challenges, we have developed a novel multi-objective optimization framework, implemented as an integrated suite of optimization methods and tools, for automatic parameter tuning of segmentation workflows in pathology image analysis and evaluated it with three real-world segmentation applications. In most experiments, we have observed significant improvements over default parameter values. Our framework was able to improve the average quality of the 15 images in 1.28× and at the same time decrease the segmentation execution time by 11.79×. The impacts of these improvements are very significant to provide segmentation results with a better quality. This should in turn allow for attaining better overall analysis results in integrated studies using cell level characterization, which are steps that typically follow the segmentation and feature extraction phases. This is essential for the enabling the use of these technologies in clinical settings, as accurately segmented objects/extracted features will lead more reliable results. Further, the gains in speedup will provide the ability of quick analyzing large-scale datasets, which are becoming available but are not yet fully exploited. Thus, we expect that pathology image analysis workflows should be submitted for a systematic tuning, such as proposed in this paper, before they are used in practice in order to maximize their benefits.

In addition, the evaluation of multiple optimization algorithms has highlighted that a single algorithm will not always be able to attain the best performance. Instead, depending on the optimization task configuration (objective, workflow choice, and input image) different algorithms may attain better performance. For instance, the BOA algorithm attained good results in single-objective runs, but was less efficient than the GA in a configuration in which we want to tune for quality and execution time.

**ACKNOWLEDGMENTS**
Omitted for blind review.